\documentclass{article}
\usepackage[utf8]{inputenc}
\usepackage{authblk}
\usepackage{setspace}
\usepackage[margin=1.25in]{geometry}
\usepackage{graphicx}
\graphicspath{ {./figures/} }
\usepackage{subcaption}
\usepackage{amsmath}
\usepackage[numbers,sort&compress]{natbib}

\title{High harmonic generation light source with polarization selectivity and sub-100-$\mu$m beam size for time- and angle-resolved photoemission spectroscopy}

\author[1]{Haoyuan Zhong}
\author[1]{Xuanxi Cai}
\author[1]{Changhua Bao}
\author[1]{Fei Wang}
\author[1]{Tianyun Lin}
\author[2]{Yudong Chen}
\author[2]{Sainan Peng}
\author[1]{Lin Tang}
\author[1]{Chen Gu}
\author[2]{Zhensheng Tao}
\author[1]{Hongyun Zhang}
\author[1,3*]{Shuyun Zhou}

\affil[1]{State Key Laboratory of Low-Dimensional Quantum Physics and Department of Physics, Tsinghua University, Beijing 100084, P. R. China.}
\affil[2]{State Key Laboratory of Surface Physics, Key Laboratory of Micro and Nano Photonic Structures (MOE), and Department of Physics, Fudan University, Shanghai 200433, P. R. China.}
\affil[3]{Frontier Science Center for Quantum Information, Beijing 100084, P. R. China.}
\affil[*]{Address correspondence to: syzhou@mail.tsinghua.edu.cn}

\date{}

\begin{document}

\maketitle

\begin{abstract}

High-quality ultrafast light sources are critical for developing advanced time- and angle-resolved photoemission spectroscopy (TrARPES). While the application of high harmonic generation (HHG) light sources in TrARPES has increased significantly  over the past decade, the optimization of the HHG probe beam size and selective control of the light polarization, which are important for TrARPES measurements, have been rarely explored. In this work, we report the implementation of high-quality HHG probe source with an optimum beam size down to 57 $\mu$m$\times$ 90 $\mu$m and selective light polarization control, together with mid-infrared (MIR) pumping source  for TrARPES measurements using a 10 kHz amplifier laser.  The selective polarization control  of the HHG probe source allows to enhance bands with different orbital contributions or symmetries, as demonstrated by experimental data measured on a few representative transition metal dichalcogenide materials (TMDCs) as well as topological insulator Bi$_2$Se$_3$. Furthermore, by combining the HHG probe source with MIR pumping at  2 $\mu$m wavelength, TrARPES on a bilayer graphene shows a time resolution of 140 fs, allowing to distinguish two different relaxation processes in graphene. Such high-quality HHG probe source together with the MIR pumping expands the capability of TrARPES in revealing the ultrafast dynamics and light-induced emerging phenomena in quantum materials.  

\end{abstract}


\section{Introduction}

Time- and angle-resolved photoemission spectroscopy (TrARPES) is a powerful experimental technique for capturing the ultrafast dynamics of quantum materials  \cite{Smallwood_2016,zxrev2021,DamascelliRMP2023,zhourev2022,na2023advancing} with full energy-, momentum-, and time-resolved information. Over the past decades, TrARPES has witnessed a rapid development in instrumentation, which not only allows to capture the ultrafast electronic dynamics of quantum materials, but also opens up new opportunities for utilizing the strong pump pulse as a control knob for tailoring the transient electronic structure \cite{BS_Floquet_2013S,zhou_2023nature_bp,zhou_2023prl_bp,bao2022light} or inducing a phase transition \cite{nonlinearrev_heieh2021,zong2021unconventional,disa2021engineering,Sentef2021RMP}.  For TrARPES instruments, the technical specifications of the pump and probe beams are critical in determining their performance. While the pump determines how light excites or couples with the materials under investigation, the probe beam determines the types of materials as well as energy and momentum range that can be probed.  Fourth harmonic generation (FHG) light source generated from solid-state nonlinear crystals provides a convenient probe light source with photon energy up to 6.3 eV based on BBO or LBO crystals \cite{bbo_perfetti,smallwood2012ultrafast,bbo_Patrick2020,Zhou2021RSI}, or a tunable photon energy of 5.3 -- 7.0 eV based on KBBF devices \cite{shin_2007KBBF,ZhouXJ2007KBBF,AK_2014KBBF,Zhou2022RSI,sz2022RSI100fs}.  While such FHG light sources have the advantages of achieving high energy and momentum resolution, it only covers $\sim$ 1 eV energy range of the photoelectrons below the Fermi energy $E_F$ (after subtracting the work function of $\sim$ 4.3 eV), with an accessible momentum range limited up to $\sim$ 0.6 \AA$^{-1}$~\cite{sz2022RSI100fs} near the Brillouin zone (BZ) center. Such energy and momentum range is not enough to probe the interesting electronic structure of many materials, e.g. the conical dispersion of graphene or the momentum valleys of transition metal dichalcogenides (TMDCs) at large momentum.

Extreme ultraviolet (XUV) light sources generated by high harmonic generation (HHG) using inert gas as nonlinear medium can provide sufficiently high photon energy to access a large energy and momentum range required for most quantum materials. Over the past decades,  HHG light sources have been increasingly applied in  TrARPES measurements  \cite{bauer_2007RSI,Weinelt_2013RSI,martin_2014JESRP,gedik_2019NC,Raphl_2019RSI,kaindl_2019RSI,jones_2019RSI,mathias_2020RSI,Chinil_2020RSI,DH_2023RSI},   making it possible to reveal the ultrafast dynamics of materials such as TMDCs \cite{bauer_2011tise2,bauer_2012tise2,exciton_2021NS} and graphene \cite{gierz2013snapshots}. While major improvement in the energy resolution and photon flux has been achieved, the typical beam size is often larger than 100 $\mu m$ \cite{Raphl_2019RSI,kaindl_2019RSI,jones_2019RSI,mathias_2020RSI,Weinelt_2013RSI}, making it unsuitable for probing small samples such as exfoliated flakes or van der Waals heterostructures. Achieving a more tightly focused beam size is important to extend the  ultrafast dynamics investigation to include those interesting samples. In addition, another highly desirable property for the probe beam is the selective polarization control for the HHG source, since different light polarizations can selectively enhance the intensity of individual bands with different orbital contributions and symmetries due to the dipole matrix element \cite{zxrev2003,CD_2020PRL_Ralph,CD_2022PRX_Ralph}.

Here, we report the implementation of a high-quality HHG light source with mid-infrared (MIR) pumping for TrARPES measurements  using a 10 kHz laser amplifier. Two HHG schemes using capillary and gas cell are compared to reveal their advantages and disadvantages. The HHG light source combines the advantages of high photon flux, an energy resolution better than 80 meV. By developing a systematic  optimization strategy for the HHG beam size, an optimized beam size down to 57 $\mu$m$\times$ 90 $\mu$m is achieved. Moreover, a polarization selectivity of the HHG light source between $s$-polarization ($s-pol.$) and $p$-polarization ($p-pol.$) is also achieved by switching the polarization of the driving laser. The power of such polarization selectivity in enhancing bands with different orbital contributions or symmetries is demonstrated by experimental results on TMDCs, such as NbSe$_2$, MoSe$_2$ and TiSe$_2$, as well as topological insulator (TI) Bi$_2$Se$_3$. Finally, the high pulse energy of the 10 kHz laser also allows MIR pumping TrARPES with  an overall time resolution of 140 fs, as demonstrated by TrARPES measurements on a bilayer graphene. Such system expands the capability of TrARPES in revealing the ultrafast dynamics and light-induced emerging phenomena in quantum materials.

\begin{figure}[h]
    \centering
    \includegraphics[width=15cm]{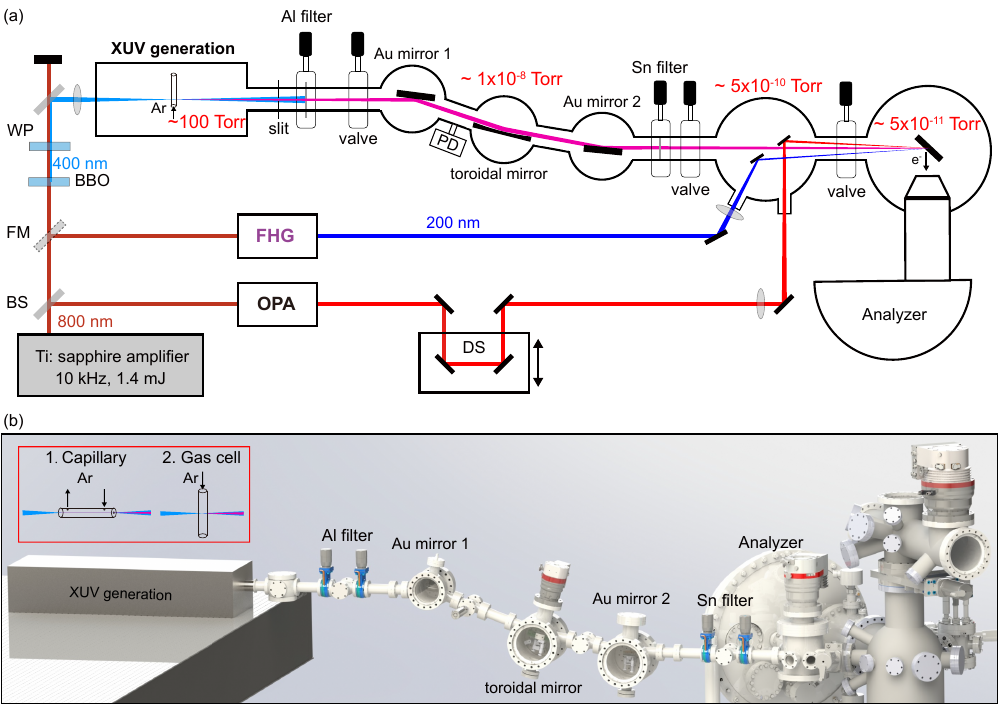}
    \caption{A schematic layout of TrARPES setup with HHG probe and MIR  pump light sources. (a) A schematic layout of the TrARPES setup, including HHG probe, forth harmonic generation (FHG) probe and MIR pump. BS: beam splitter; FM: flip mirror; WP: waveplate; DS: delay stage. The incident angle is 80$^\circ$, 85$^\circ$ and 85$^\circ$ for gold mirror 1, toroidal mirror and gold mirror 2. (b) Engineering drawing of the HHG beamline. The inset shows two HHG setups using capillary and gas cell, respectively.}
    \label{Fig1}
\end{figure}

\section{DESIGNING PRINCIPLES AND EXPERIMENTAL SETUP}

Figures 1 shows an overview of the TrARPES system. Near-infrared pulses at 800 nm with a pulse energy of 1.4 mJ and pulse duration of 35 fs are generated by a Ti:Sapphire amplifier at 10 kHz repetition rate.  Thanks to the high pulse energy, our TrARPES system can have versatile pump and probe beams through nonlinear optical processes, for example, MIR pumping, FHG and HHG probes. The MIR pumping is useful for generating a stronger light-matter interaction which is critical for Floquet engineering \cite{BS_Floquet_2013S,zhou_2023nature_bp,zhou_2023prl_bp}. The two types of probe light sources are complementary to each other: while the FHG source has higher resolution for zooming-in the fine band structure near the BZ center, the HHG source can probe a sufficiently large energy and momentum range to cover the entire BZ of most materials. Such combination can maximize the performance of the TrARPES system.

In this work, we focus mainly on the implementation of a high-quality HHG light source and its optimization. Below we discuss the main designing principles of our HHG light source.  In order to achieve higher efficiency and easier isolation of the desired harmonic, the HHG light source is driven by the second harmonic (SH) of the fundamental beam \cite{martin_2014JESRP,kaindl_2019RSI}. Before entering the main analysis chamber, the HHG beam is steered by Au mirrors, focused by a toroidal mirror and filtered by metallic films as shown in Fig.~1(a). The metallic films have three functions: blocking the residual SH; isolating desired single harmonics (7th order by Al and Sn \cite{martin_2014JESRP,kaindl_2019RSI,Raphl_2019RSI, Chinil_2020RSI}); and protecting high vacuum together with a 0.5 mm slit. A high-quality HHG source needs to have desired properties, such as high photon flux, good stability, high energy resolution, small beam size, polarization selectivity etc. In the following, we introduce how the HHG light source is designed to satisfy these important requirements.

\begin{figure}[htbp]
    \centering
    \includegraphics[width=15cm]{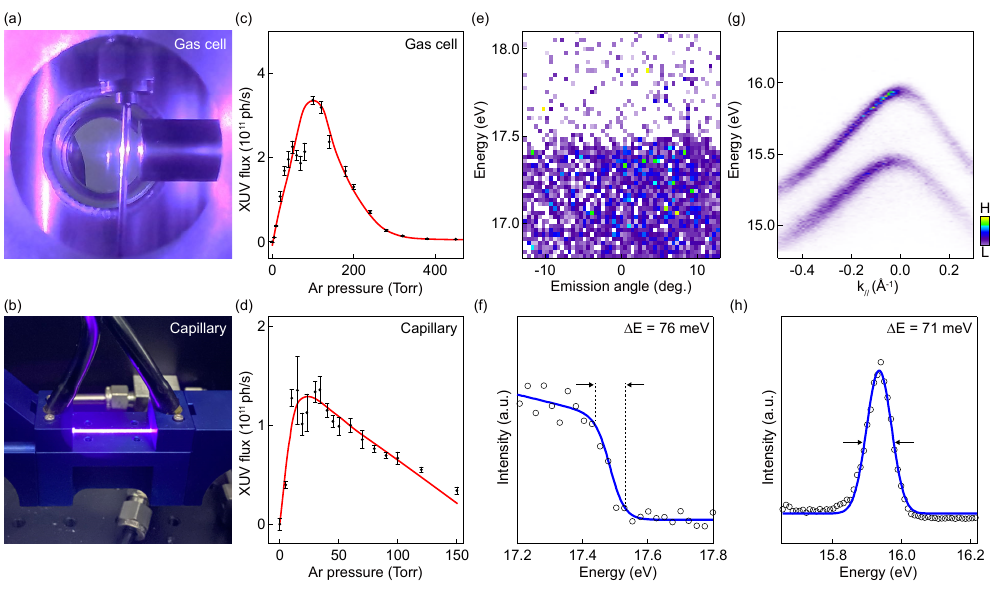}
    \caption{Characterization of the HHG source. (a) and (b) Photos of gas cell and  capillary during operation. (c, d) Photon flux of the HHG source as a function of Ar pressure  generated by (c)gas cell and (d) capillary, where the input SH is 150 $\mu$J and the  generated HHG source passes through a 0.2 $\mu$m thick Al film and a Au mirror (grazing incidence). (e) Photoemission spectrum of polycrystalline gold measured by HHG light generated by gas cell, which shows a clear Fermi edge. (f) Energy distribution  curve (EDC) extracted from (e), which is fitted by a Fermi-Dirac distribution. The photon energy is fitted to be 21.78 eV (17.48 eV Fermi energy plus a 4.3 eV work function). The total energy resolution is 76 $\pm$ 20 meV after subtracting thermal broadening of 28 meV at the measurement temperature of 80 K and an analyzer broadening of 40 meV. (g) Electronic structure of bulk WSe$_2$ at K point, which shows sharp spectrum. (h) EDC extracted from (g), which is fitted by Gaussian function. The total energy energy resolution is 71 $\pm$ 2 meV after subtracting thermal broadening at 80 K and an analyzer broadening .}
    \label{Fig2}
\end{figure}

\section{Results}

Here, we report the experimental design and characterizations of the HHG source, in terms of the photon flux, energy resolution and beam size. More importantly, the high performance TrARPES system  with polarization selectivity and high time resolution are further demonstrated by ARPES and TrARPES measurements on TMDCs, TI as well as graphene. 

\subsection{Generation and characterizations of the high-quality HHG source}

We first evaluate two approaches for generating HHG by using a gas cell or a capillary. The gas cell is a 1-mm inner diameter nickel tube with  {\it in situ} drilled holes for laser transmission as shown in Fig.~2(a), and the capillary is a 6 cm-long, 150 $\mu$m-inner-diameter quartz waveguide as shown in Fig.~2(b). These two approaches have their own advantages and disadvantages. Although the capillary approach can in principle generate higher harmonics due to the waveguide confinement \cite{HHGcap_1998Sci_KM,HHGcap_1999PRL_KM}, it also places more stringent requirements on the input pulse energy (often higher than 1 mJ \cite{bauer_2007RSI,martin_2014JESRP,HHG_Kai_2022RSI}) and input beam (alignment, size and mode) in order to achieve better confinement in the waveguide. In contrast, the gas cell approach is much more tolerant to the input pulse energy and beam (since there is less beam  confinement), where a pulse energy of 10 $\mu$J is still feasible if the beam is focused tightly enough \cite{tight_focus_2016Optica,mathias_2020RSI}. Secondly, the stability of the HHG source is better by using gas cell approach (note that beam-pointing  stabilization is not applied for both approaches), with a root-mean-square (RMS) fluctuation of 5$\%$ compared to 20$\%$ for the capillary as shown in Figs.~ 2(c,d).  The above comparison suggests that the gas cell is a better choice for our applications, with less stringent beam requirements and better stability.

Figure 2(c) shows the photon flux of the HHG source as a function of Ar pressure using a gas cell. The phase-matching condition for HHG is attained when the Ar gas pressure approaches 100 Torr for the gas cell, which results in a maximum photon flux of 3.5$\times$10$^{11}$ photons/s measured by a photodiode with diode responsibility of 0.25 A/W. The energy resolution of the HHG source is extracted by the Fermi edge measured on a polycrystalline Au as shown in Fig.~2(e), where the angle-integrated curve is fitted by Fermi-Dirac distribution function in Fig.~2(f). The energy resolution contributed by the probe source is 76 $\pm$ 20 meV. The high energy resolution is also reflected in the sharp dispersion image measured on a single crystal WSe$_2$ in Fig.~2(g), from which an analysis of the integrated intensity in Fig.~2(h) shows an overall linewidth of 71 $\pm$ 2 meV, consistent with the energy resolution of the probe source. The high-quality data demonstrate that our TrARPES has a probe energy resolution better than 80 meV, which is sufficient for high-resolution ARPES experiments.

\subsection{Optimization of the HHG beam size}

Achieving a tightly focused HHG beam better than 100 $\mu$m is critical for expanding the probing capability of TrARPES system to small samples, such as exfoliated flakes and van der Waals heterostructures. The HHG beam is focused using a toroidal mirror with a 1:1 imaging ratio: the focal length of the toroidal mirror is 750 mm, by which the diverging HHG beam from the source is refocused onto the sample. Since the HHG source requires vacuum condition to propagate and is complicated to align, it is more convenient to use a laser pointer in the atmosphere to pre-align the toroidal mirror and to develop a strategy for the beam optimization, given the fact that the toroidal mirror is a reflective optics and is in principle wavelength-independent. 

\begin{figure}[htbp]
    \centering
    \includegraphics[width=15cm]{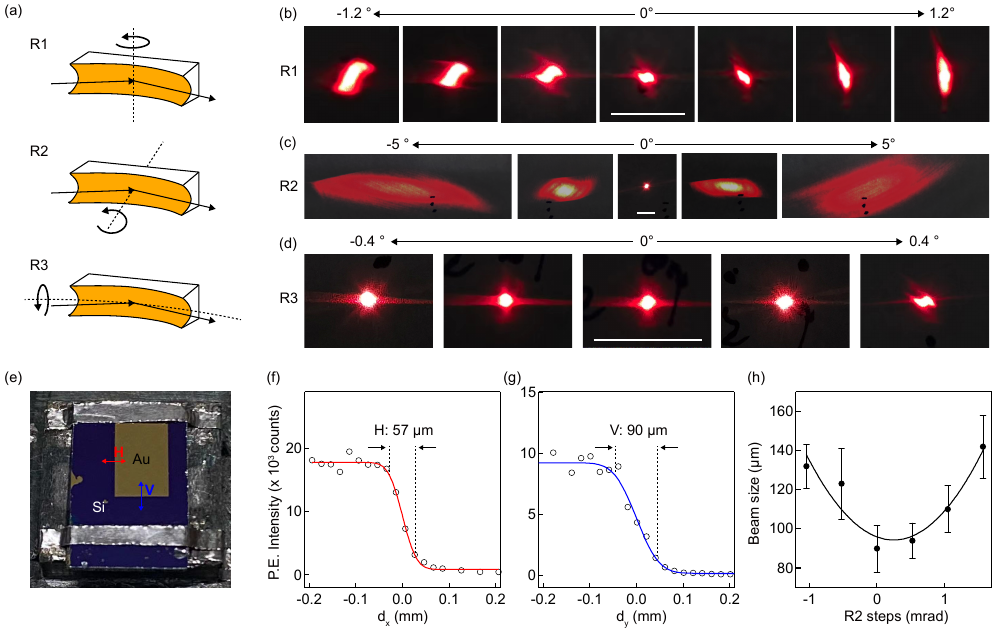}
    \caption{Optimization of beam focusing using toroidal mirror. (a) The schematic of a toroidal mirror, which processes three degrees of freedom: R1, R2 and R3 rotation. (b--d)  Photos of the laser pointer beam at focal point as  function of R1--R3 rotations (the white lines are uniform scale bars for R1--R3 which represent the same length). (e) Photo of Au pattern (thickness is 100 nm) deposited on insulating SiO$_2$ which is used to scan the HHG beam size in two directions. The edge of the Au pattern is sharper than 1 $\mu$m. (f) and (g) The photoemission intensity obtained from horizontal (red) and vertical (blue) line cuts as marked in (e). The intensity contrast is fitted by convolution of a step function and a Gaussian function. (h) Vertical beam size as a function of R2 rotation. }
    \label{Fig3}
\end{figure}

There are three degrees of rotation freedom for the toroidal mirror as shown in Fig.~3(a): yaw rotation R1, pitch rotation R2 and roll rotation R3. Figures 3(b--d) show the images of the focused beam obtained by rotating R1--R3, indicating that the beam size is sensitive to R1 and R2, while not sensitive in current range of R3. Following such principle, we optimize the toroidal mirror by achieving a smallest beam size of driving laser at the sample position. Finally, to further calibrate the HHG beam size in the UHV directly, we further use a Au pattern on insulating SiO$_2$ (edges sharper than 1 $\mu$m)  in Fig.~3(e) for knife-edge scans.  By moving the Au pattern and scanning the ARPES intensity across the horizontal or vertical edges (Figs.~3(f,g)), the optimized beam size is measured to be 57 $\mu$m $\times$ 90 $\mu$m.  We notice that the vertical beam size is quite sensitive to the R2 rotation.  Figure 3(h) shows the vertical beam size as a function of R2 steps, where the beam size almost doubles from 0 to -1 mrad  rotation, which demonstrates the HHG beam size is extremely sensitive to R2 rotation and should be carefully adjusted. Such small beam size is possible after carefully optimizing the toroidal mirror using the above strategy, which paves an important step toward measuring small samples (e.g. exfoliated flakes of graphene, TMDCs and their heterostructures) and obtaining high-quality ARPES spectra.

\subsection{High performance ARPES measurements with polarization selectivity}

\begin{figure}[h]
    \centering
    \includegraphics[width=15cm]{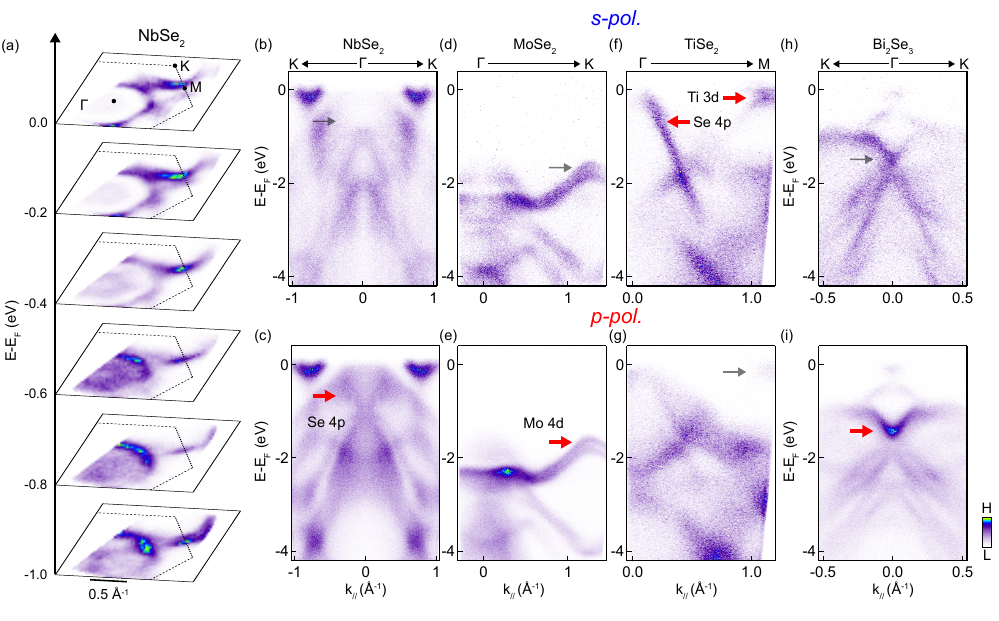}
    \caption{Electronic structure of various quantum materials measured by the HHG source. (a) Energy contours of NbSe$_2$. (b)--(i) Experimental electronic structure of various various TMDCs including NbSe$_2$, MoSe$_2$, TiSe$_2$, and topological insulator Bi$_2$Se$_3$ using $s-pol.$ (top rows) and $p-pol.$ (bottom rows) HHG polarizations respectively.  The measurement direction is along $\Gamma$-K for (b,d,h) and $\Gamma$-M direction for (f).}
    \label{Fig4}
\end{figure} 

In addition to the good energy resolution and optimized beam size discussed above, the implemented HHG source also has polarization selectivity.  We note that polarization control of the HHG source is important  for ARPES measurements, because bands with different orbitals can have different dipole matrix elements. Therefore, in order to  have a non-zero or a stronger photoemission intensity for different orbitals, different light polarizations are required to align the electrical field of the incident light parallel to orbitals \cite{zxrev2003}. Moreover, such dipole matrix element using different light polarizations also carries useful information about the orbital pseudospin and Berry curvature of the materials \cite{kim_BP_pseudospin2020,Kekule_2021PRL,CD_2020PRL_Ralph,CD_2022PRX_Ralph, atto_2016Sci_Tao,atto_2020JPB_Shi}. 

In our experimental setup,  the polarization of the output HHG source can be easily controlled by a wave plate (WP), which is inserted before the HHG chamber to control the polarization of the driving laser.  Although the polarization selectivity between $s-pol.$ and  $p-pol.$ by switching the polarization of the driving laser has been discussed \cite{kaindl_2019RSI}, how the HHG  polarization affects ARPES measurements has been much less explored \cite{atto_2016Sci_Tao, CD_2022PRX_Ralph}. 
The developed HHG light source shows high performance, which are demonstrated by ARPES measurements on various TMDCs materials such as NbSe$_2$, MoSe$_2$, TiSe$_2$, as well as topological insulator Bi$_2$Se$_3$. Figure~4(a) shows three-dimensional electronic structure measured on NbSe$_2$, and Figs.~4(b,c) shows a comparison of dispersion images measured along the K-$\Gamma$-K direction by using $s-pol.$ and $p-pol.$ HHG polarizations. It is clear that bands contributed by Se $4p$ orbitals (pointed by black arrow in Figs.~4(b,c)) \cite{NSARPES_2022NP,NSARPES_2023PRM} show a much stronger intensity by using $p-pol.$ HHG polarization. Similarly, for MoSe$_2$, the spin-orbital splitting of the Mo $4d$ bands \cite{MSARPES_2012PRM,MSARPES_2014NN} near the K valley (marked by black arrow in Figs.~4(d,e)) are better resolved when switching from $s-pol.$ (Fig.~4(d)) to $p-pol.$ probe (Fig.~4(e)). We note that although $p-pol.$ tends to give a stronger intensity for the $d$ and $p$ orbital of TMDCs, it is not always better for resolving the bands, especially when the sample geometry changes.  For example, when the sample geometry changes to $\Gamma$-M direction (90$^{\circ}$-rotation compared to $\Gamma$-K) in charge density wave (CDW) material TiSe$_2$ \cite{TSARPES_2011Na}, the folded band from Se $4p$ orbitals, which is induced by CDW order and Ti $3d$ bands at the M point (marked black arrow) of TiSe$_2$, are better revealed using $s-pol.$ probe (Fig.~4(f)) than  $p-pol.$ probe (Fig.~4(g)). Besides TMDCs mentioned above, suitable polarization is also important for measuring Bi$_2$Se$_3$ as shown in Figs.~4(h,i). These examples demonstrate the power of the HHG polarization selectivity in obtaining high-quality data quality  when measuring different materials or geometries.

\subsection{High resolution TrARPES with HHG probe and MIR pumping}

The high-quality HHG probe source can be further integrated with MIR pumping, allowing to excite the sample more selectively, to distinguish different relaxation processes, and further to manipulate the transient electronic structure via Floquet engineering. Below we use graphene as an example to demonstrate the TrARPES performance with MIR pumping. 

Figure~5(a) shows Dirac cone band structure of epitaxy bilayer graphene on SiC, which shows two sets of valence band with slight electron doping \cite{blgarpes_2006Sci}. Figure 5(b)--(f) show TrARPES snapshots of the Dirac cone at different delay times upon pumping at 2 $\mu$m wavelength. The corresponding pump photon energy is 540 meV, and electrons are photo-excited into the unoccupied conduction bands above E$_F$ as shown in Fig.~5(b). The hot electrons then relax through electron-phonon coupling.

In order to track the ultrafast dynamics of the photo-exited electrons continuously, Fig.~5(g) shows the evolution of momentum-integrated intensity as a function of energy and delay time, which can provide information about time resolution and thermodynamics. First, the time resolution can be extracted by analyzing the time trace at energy 0.5 eV, which is near direct optical excitation. The time trace is fitted by a Gaussian function, from which the instrumental time resolution is extracted to be 140 $\pm$ 10 fs. Second, the photo-excited thermodynamics can also be revealed by tracing the evolution of the transient electronic temperature $T_e$ by fitting the energy distribution curves (EDCs) with a Fermi-Dirac distribution as shown in Fig.~5(i). The transient electronic temperature is fitted by the product of the step function and a double-exponential function convolved with a Gaussian function as shown in Fig.~5(j) \cite{szhoukekule2022nsr}. The fast decay time $\tau_1$ = 90 $\pm$ 10 fs corresponds to decay of hot electrons via optical phonons, while the slow decay $\tau_2$ = 1800 $\pm$ 200 fs corresponds to decay of hot electrons via acoustic phonons, in agreement with previous reports \cite{BLG_TrARPES_2014PRL,BLG_TrARPES_2015jpcm}.

\begin{figure}[h]
    \centering
    \includegraphics[width=15cm]{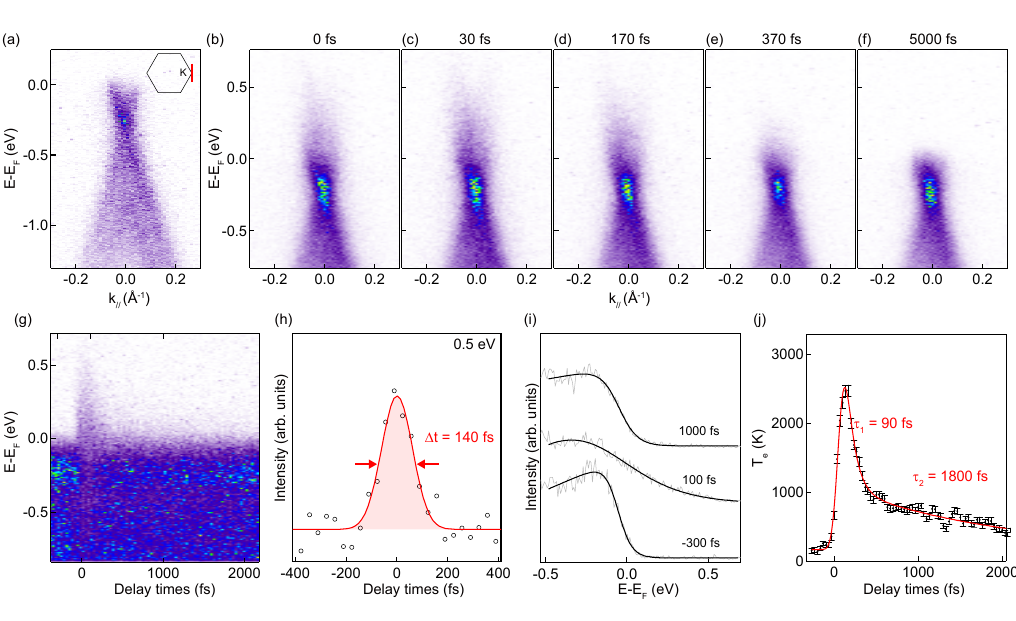}
    \caption{Ultrafast dynamics of bilayer graphene revealed by TrARPES upon 2 $\mu$m pumping. (a) Dispersion image of epitaxy bilayer graphene at K point measured by $p-pol.$ HHG photons. Inset: Brillouin zone with measurement direction (red line). (b)--(f) TrARPES dispersion images of bilayer graphene at different delay times. The pump photon energy is 540 meV and the pump fluence is 1 mJ cm$^{-2}$ with $p-pol.$ (g) Evolution of momentum-integrated intensity with energy and delay times. (h) TrARPES intensity as a function of delay times at 0.5 eV. The red curve is the fitted Gaussian function for extracting the time resolution. (i) EDCs at different delay times from (g) and corresponding Fermi-Dirac fitting. (j) Extracted electronic temperature as a function of delay time.}
    \label{Fig4}
\end{figure} 

\section{Discussions}

In summary, we have successfully implemented a HHG light source for high-performance TrARPES measurements. The HHG light source has a high photon flux, with an energy resolution better than 80 meV, a small beam size of 57 $\mu m$ $\times$ 90 $\mu$m after careful optimization of the focusing toroidal mirror and a high time resolution of 140 fs, which makes it applicable for revealing the ultrafast dynamics and light-induced phenomena of quantum materials, including exfoliated flakes and van der Waals heterostructures. More importantly, high performance of the HHG light source in ARPES measurements and the power of polarization selectivity have been demonstrated by ARPES measurements on a few representative TMDCs and TI, emphasizing the competent capacity of the HHG light as well as the importance of polarization control to access band structure when measuring different materials. The compatibility of HHG probe with 2 $\mu$m MIR pumping is also demonstrated by TrARPES measurements on graphene, from which a time resolution of 140 fs is achieved. 

Finally, we would like to comment on the implications of generating HHG source using an amplifier laser operating in a repetition rate of  10 kHz and high pulse energy of 1.4 mJ. Such system combines the advantage of high-quality data acquisition (ARPES measurements in Fig.~4) and the implementation of a MIR pump source, which is necessary for Floquet engineering. We note that so far TrARPES investigations of Floquet engineering have been focused only on a few semiconductors \cite{WSe2_Isabella2021,zhou_2023nature_bp,zhou_2023prl_bp} and topological insulators \cite{BS_Floquet_2013S,ito2023build}. Here our work shows that detection of Floquet states at the K point of graphene or valley selective Floquet engineering of TMDCs would become feasible when combing with MIR pumping. 

\section*{Acknowledgments}

\subsection*{General} 
We thank Xiaoshi~ Zhang and Chenxia~ Yun for useful discussions. 

\subsection*{Author Contributions} 
S.Z. conceived the research project. H. Zhong, X.C. and C.B. design and build up the HHG light source. H. Zhong, X.C., F.W. and T.L. performed the ARPES measurements and analysed the date. W.F. grew the single crystals. H. Zhong and S.Z. wrote the manuscript, and all authors contributed to the discussions and commented on the manuscript.
\subsection*{Funding}
This work is supported by National Natural Science Foundation of China (Grant No.~12327805, 12234011, 92250305, 52388201, 11725418 and 11427903), and the National Key R$\&$D Program of China (No. 2021YFA1400100, 2020YFA0308800). C.B. is supported by a project funded by China Postdoctoral Science Foundation (No.~2022M721886 and No.~BX20230187) and the Shuimu Tsinghua Scholar Program. Z. T. acknowledges the support from the National Key Research and Development Program of China (Grant No.~2021YFA1400200), the National Natural Science Foundation of China (No.~12274091) and the Shanghai Municipal Science and Technology Basic Research Project (Grant No.~22JC1400200).

\subsection*{Competing interests}
The authors declare that there is no conflict of interest regarding the publication of this article.

\subsection*{Data Availability}
The data presented in this study are available on reasonable request from the corresponding author.

\end{document}